\documentstyle[preprint,aps]{revtex}
\input epsf.sty
\begin{document}
\draft
\title
{Quantum Computation with Quantum Dots}
\author{Daniel Loss$^{a,b}$\footnote{loss@ubaclu.unibas.ch} and David P.
DiVincenzo$^{a,c}$\footnote{divince@watson.ibm.com}}
\address{$^a$Institute for Theoretical Physics, University of
California,
Santa Barbara, CA 93106-4030, USA\\
$^b$Department of Physics and Astronomy, University of Basel,
Klingelbergstrasse 82, 4056 Basel, Switzerland\\
$^c$IBM Research Division, T. J. Watson Research Center,
P. O. Box 218, Yorktown Heights, NY 10598, USA\\
}
\date{\today}
\maketitle
\begin{abstract}
We propose a new implementation of a universal set of one- and
two-qubit gates for quantum computation using the spin states of
coupled single-electron quantum dots.  Desired operations are effected
by the gating of the tunneling barrier between neighboring dots.
Several measures of the gate quality are computed within a newly
derived spin master equation incorporating decoherence caused by a
prototypical magnetic environment.  Dot-array experiments which would
provide an initial demonstration of the desired non-equilibrium spin
dynamics are proposed.
\end{abstract}

\section{Introduction}

The work of the last several years has greatly clarified both the
theoretical potential and the experimental challenges of quantum
computation\cite{BD}.  In a quantum computer the state of each bit is
permitted to be any quantum-mechanical state of a {\em qubit}
(two-level quantum system).  Computation proceeds by a succession of
``two-qubit quantum gates''\cite{G9}, coherent interactions involving
specific pairs of qubits, by analogy to the realization of ordinary
digital computation as a succession of boolean logic gates.  It is now
understood that the time evolution of an arbitrary quantum state is
intrinsically more powerful computationally than the evolution of a
digital logic state (the quantum computation can be viewed as a
coherent superposition of digital computations proceeding in
parallel).

Shor has shown\cite{Shor} how this parallelism may be exploited to
develop polynomial-time quantum algorithms for computational problems,
such as prime factoring, which have previously been viewed as
intractable.  This has sparked investigations into the feasibility of
the actual physical implementation of quantum computation.  Achieving
the conditions for quantum computation is extremely demanding,
requiring precision control of Hamiltonian operations on well-defined
two-level quantum systems, and a very high degree of quantum
coherence\cite{DDV}.  In ion-trap systems\cite{CZ}, and in cavity
quantum electrodynamic experiments\cite{JK}, quantum computation at
the level of an individual two-qubit gate has been demonstrated;
however, it is unclear whether such atomic-physics implementations
could ever be scaled up to do truly large-scale quantum computation,
and some have speculated that solid-state physics, the scientific
mainstay of digital computation, would ultimately provide a suitable
arena for quantum computation as well.  The initial realization of the
model that we introduce here would correspond to only a modest step
towards the realization of quantum computing, but it would at the same
time be a very ambitious advance in the study of controlled
non-equilibrium spin dynamics of magnetic nanosystems, and could point
the way towards more extensive studies to explore the large-scale
quantum dynamics envisioned for a quantum computer.

\section{Quantum-dot implementation of two-qubit gates}

In this report we develop a detailed scenario for how quantum
computation may be achieved in a coupled quantum dot
system\cite{prev}.  In our model the qubit is realized as the spin of
the excess electron on a single-electron quantum dot, see Fig. 1.  We
introduce here a novel mechanism for two-qubit quantum-gate operation
that operates by a purely electrical gating of the tunneling barrier
between neighboring quantum dots, rather than by spectroscopic
manipulation as in other models.  Controlled gating of the tunneling
barrier between neighboring single-electron quantum dots in patterned
two-dimensional electron-gas structures has already been achieved
experimentally using a split-gate technique\cite{Harvard}.  If the
barrier potential is ``high'', tunneling is forbidden between dots,
and the qubit states are held stably without evolution in time ($t$).
If the barrier is pulsed to a ``low'' voltage, the usual physics of
the Hubbard model\cite{AM} says that the spins will be subject to a
transient Heisenberg coupling,
\begin{equation}
H_s(t)=J(t) \vec S_1\cdot\vec S_2\,\, ,\label{tham}
\end{equation}
where $J(t)=\frac{4 t_0^2(t)}{u}$ is the time-dependent exchange
constant\cite{superexchange} which is produced by the turning on and
off of the tunneling matrix element $t_0(t)$.  Here $u$ is the
charging energy of a single dot, and $\vec S_i$ is the spin-1/2
operator for dot $i$.

Eq.\ (\ref{tham}) will provide a good description of the quantum-dot
system if several conditions are met: 1) Higher-lying single particle
states of the dots can be ignored; this requires $\Delta E\gg kT$,
where $\Delta E$ is the level spacing and $T$ is the temperature.  2)
The time scale $\tau_s$ for pulsing the gate potential ``low'' should
be longer than $\hbar/\Delta E$, in order to prevent transitions to
higher orbital levels.  3) $u>t_0(t)$ for all $t$; this is required
for the Heisenberg-exchange approximation to be accurate.  4) The
decoherence time $\Gamma^{-1}$ should be much longer than the
switching time $\tau_s$.  Much of the remainder of the paper will be
devoted to a detailed analysis of the effect of a decohering
environment.  We expect that the spin-1/2 degrees of freedom in
quantum dots should generically have longer decoherence times than
charge degrees of freedom, since they are insensitive to any
environmental fluctuations of the electric potential.  However, while
charge transport in such coupled quantum dots has received much recent
attention\cite{Glazman,Harvard} we are not aware of investigations on
their non-equilibrium spin dynamics as envisaged here.  Thus, we will
carefully consider the effect of magnetic coupling to the environment.

If $\Gamma^{-1}$ is long, then the ideal of quantum computing may be
achieved, wherein the effect of the pulsed Hamiltonian is to apply a
particular unitary time evolution operator $U_s(t)=T\exp\{-i\int_0^t
H_s(t')dt'\}$ to the initial state of the two spins:
$|\Psi(t)\rangle=U_s|\Psi(0)\rangle$.  The pulsed Heisenberg coupling
leads to a special form for $U_s$: For a specific duration $\tau_s$ of
the spin-spin coupling such that $\int dt J(t)=J_0\tau_s=\pi$ (mod
$2\pi$)\cite{pulseform}, $U_s(J_0\tau_s=\pi)=U_{sw}$ is the ``swap''
operator: if $|ij\rangle$ labels the basis states of two spins in the
$S_z$-basis with $i,j=0,1$, then $U_{sw}|ij\rangle=|ji\rangle$.
Because it conserves the total angular momentum of the system,
$U_{sw}$ is not by itself sufficient to perform useful quantum
computations, but if the interaction is pulsed on for just half the
duration, the resulting ``square root of swap'' is very useful as a
fundamental quantum gate: for instance, a quantum XOR gate is obtained
by a simple sequence of operations:
\begin{equation}
U_{XOR}=
e^{i {\pi\over 2}S_1^z }e^{-i {\pi\over 2}S_2^z}
U_{sw}^\frac{1}{2}e^{i\pi S_1^z}U_{sw}^\frac{1}{2}\,\, ,
\label{XOR}
\end{equation}
where $e^{i\pi S_1^z}$ etc. are single-qubit operations only, which
can be realized e.g. by applying local magnetic fields (see
Sec. \ref{1b}) \cite{footnoteHXOR}.  It has been established that XOR
along with single-qubit operations may be assembled to do any quantum
computation\cite{G9}.  Note that the XOR of Eq.\ (\ref{XOR}) is given
in the basis where it has the form of a conditional phase-shift
operation; the standard XOR is obtained by a simple basis change for
qubit 2\cite{G9}.

\section{The master equation}

We will now consider in detail the {\em non-ideal} action of the swap
operation when the two spins are coupled to a magnetic environment.  A
new master equation model is obtained that explicitly accounts for the
action of the environment during switching --- to our knowledge, the
first treatment of this effect.  We use a Caldeira-Leggett type model
in which a set of harmonic oscillators are coupled linearly to the
system spins by $H_{int}=\lambda\sum_{i=1,2}\vec S_i\cdot \vec
b_i$. Here, $b_i^j=\sum_\alpha g^{ij}_\alpha(a_{\alpha,ij}+
a_{\alpha,ij}^\dagger)$ is a fluctuating quantum field whose free
motion is governed by the harmonic oscillator Hamiltonian
$H_B=\sum\omega_\alpha^{ij}a_{\alpha,ij}^\dagger a_{\alpha,ij}$, where
$a_{\alpha,ij}^\dagger$/$a_{\alpha,ij}$ are bosonic
creation/annihilation operators (with $j=x,y,z$), and
$\omega_\alpha^{ij}$ are the corresponding frequencies with spectral
distribution function $J_{ij}(\omega)=
\pi\sum_{\alpha}(g^{ij}_\alpha)^2
\delta(\omega-\omega_\alpha)$\cite{Chuang}.  The system and
environment are initially uncorrelated with the latter in thermal
equilibrium described by the canonical density matrix $\rho_B$ with
temperature $T$.  We assume for simplicity that the environment acts
isotropically and is equal and independent on both dots.  We do not
consider this to be a microscopically accurate model for these
as-yet-unconstructed quantum dot systems, but rather as a generic
phenomenological description of the environment of a spin, which will
permit us to explore the complete time dependence of the gate action
on the single coupling constant $\lambda$ and the controlled
parameters of $H_s(t)$\cite{footnoteSchoen}.

\subsection{Swap gate}

The quantity of interest is the system density matrix $\rho(t)=Tr_B
{\bar \rho} (t)$ which we obtain by tracing out the environment
degrees of freedom. The full density matrix ${\bar\rho}$ itself obeys
the von Neumann equation,
\begin{equation}
\dot{\bar \rho}(t)=-i[H,{\bar \rho}]\equiv -i{\cal L} {\bar \rho},
\label{vne}
\end{equation}
where
\begin{equation}
{\cal L}={\cal L}_s(t)+{\cal L}_{int}+{\cal L}_B \label{lio}
\end{equation}
denotes the Liouvillian\cite{Fick} corresponding to the full
Hamiltonian
\begin{equation}
H=H_s(t)+H_{int}+H_B.\label{ham}
\end{equation}
Our goal is to find the linear map (superoperator) ${\cal V}(t)$ which
connects the input state of the gate $\rho_0=\rho(t=0)$ with the
output state $\rho(t)$ after time $t>\tau_s$ has elapsed,
$\rho(t)={\cal V}(t)\,\rho_0 \, .\ $ ${\cal V}(t)$ must satisfy three
physical conditions: 1) trace preservation $Tr_s\, {\cal V}\rho=1$,
where $Tr_s$ denotes the system trace, 2) Hermiticity preservation
$({\cal V}\rho)^\dagger={\cal V}\rho$, and 3) complete positivity,
$({\cal V}\otimes 1_B)\bar\rho\ge 0$.  Using the Zwanzig master
equation approach \cite{Fick} we sketch the derivation for ${\cal V}$
in the Born and Markov approximation which respects these three
conditions.  The novel situation we analyze here is unusual in that
$H_s$ is explicitly time-dependent and changes abruptly in time. It is
this fact that requires a separate treatment for times $t\leq \tau_s$
and $t>\tau_s$. To implement this time scale separation and to
preserve positivity it is best to start from the exact master equation
in pure integral form,
\begin{equation}
\rho(t)={\cal U}_s(t,0)\rho_0
-\! \int_0^t \! d\sigma \int_0^\sigma \! d\tau {\cal U}_s(t,\sigma) {\cal M}
(\sigma,\tau)\rho(\tau),
\label{master}
\end{equation}
where
\begin{equation}
{\cal U}_i(t,t')=T \exp\{-i\int_{t'}^t d\tau {\cal
L}_i(\tau)\},\label{uuu}
\end{equation}
where $i=s,\ B,\ int,\ {\rm or}\ q$.  Here $q$ indicates the projected
Liouvillian
\begin{equation}
{\cal L}_q=(1-P){\cal L}=(1-\rho_BTr_B){\cal L}.
\end{equation}
Also, the ``memory kernel" is
\begin{equation}
{\cal M} (\sigma,\tau) =Tr_B {\cal L}_{int} {\cal
U}_q(\sigma,\tau){\cal L}_{int}\rho_B.
\end{equation}
We solve (\ref{master}) in the Born approximation and for $t\gg
\tau_s$.  To this end the time integrals are split up into three
parts, 1) $0\leq\tau\leq\sigma\leq \tau_s<t$, 2)
$0\leq\tau\leq\tau_s\leq\sigma<t$, and 3)
$0\leq\tau_s\leq\tau\leq\sigma<t$. Keeping only leading terms in
$\tau_s$ we retain the contribution from interval 2) as it is
proportional to $\tau_s$, whereas we can drop interval 1) which leads
to higher-order terms.  But note that terms containing $J_0\tau_s$
must be kept to all orders \cite{pulseform}.  Interval 3) is
independent of $\tau_s$.

Rewriting the expressions and performing a Born approximation (i.e.,
keeping only lowest order terms in $\lambda^2$)
with subsequent Markov approximation we find, for $t\geq\tau_s$,
\begin{equation}
{\cal V}(t)=e^{-(t-\tau_s){\cal K}_3}{\cal U}_s(\tau_s)\, (1-{\cal K}_2)\,\, ,
\label{k2}
\end{equation}
where ${\cal U}_i(\tau_s)={\cal U}_i(\tau_s,0)$, ${\cal K}_2$
describes the effect of the environment during the switching,
\begin{eqnarray}
{\cal K}_2=&{\cal U}&_s^\dagger(\tau_s)\int_0^{\tau_s}d\tau\int_0^\infty dt\
\nonumber\\
&\times&Tr_B {\cal L}_{int} {\cal U}_s(\tau){\cal U}_B(t)
{\cal L}_{int}\rho_B {\cal U}_s(\tau_s-\tau),\label{k2lio}
\end{eqnarray}
while
\begin{equation}
{\cal K}_3=\int_0^\infty dt\ Tr_B\ {\cal L}_{int}
{\cal U}_B(t){\cal L}_{int}\rho_B\label{k3lio}
\end{equation}
is independent of $H_s$.  We also note that ${\cal U}_s(1-{\cal K}_2)$
has a simple interpretation as being the ``transient contribution"
which changes the initial value $\rho_0$ at $t=0$ to ${\cal
U}_s(\tau_s)(1-{\cal K}_2)\rho_0$ at $t=\tau_s$.  We show in the
Appendix that, to leading order, our superoperator $\cal V$ indeed
satisfies all three conditions stated above, in particular complete
positivity.  Such a proof for
spins with an explicit time-dependent and direct interaction, Eq.
(\ref{tham}), is not simply related to the case of a master equation
for non-interacting spins (and without explicit time-dependence)
considered in the literature (see for example \cite{Davies,Fick}). We
also note that the above Born and Markov approximations could also be
introduced in the master equation in the more usual
differential-integral representation. However, it is well-known from
studies in non-interacting spin problems\cite{Celio} that in this case
the resulting propagator is in general {\it no longer} completely
positive.

Next, we evaluate the above superoperators more explicitly, obtaining
\begin{eqnarray}
{\cal K}_2\rho&=&(\Gamma+i \Delta)\sum_i \int_0^{\tau_s} d\tau
[\vec S_i(\tau_s),
\cdot\vec S_i (\tau)\rho] +h.c.  \,\,
\label{ktwo}
,\\
{\cal K}_3\rho&=&\Gamma (3\rho-2\sum_i \vec S_i\rho \cdot \vec S_i)\,\, ,
\label{kthree}
\end{eqnarray}
where $\Gamma,\Delta$ are real and given by
\begin{eqnarray}
\Gamma&=&{\lambda^2\over \pi} \int_0^\infty dt \int_0^\infty d\omega
J(\omega) \cos
(\omega t) \coth({\omega\over 2k_BT})\,  ,\\
\Delta&=&{\lambda^2\over \pi} \int_0^\infty dt \int_0^\infty d\omega
J(\omega) \sin
(\omega t)\, .
\end{eqnarray}
In our model, the transverse and longitudinal relaxation or
decoherence rates of the system spins are the same and given by
$\Gamma$. For instance, for ohmic damping with $J(\omega)=\eta
\omega$, we get $\Gamma=\lambda^2\eta k_BT$ and
$\Delta=\lambda^2\eta\omega_c/\pi$, with $\omega_c$ some high
frequency cut-off. Requiring for consistency that $\Gamma\tau_s,
\Delta\tau_s\ll 1$ we find that ${\cal K}_2$ is in fact a small
correction.  However, we emphasize again that, to our knowledge, this
is the first time that any analysis of this ${\cal K}_2$ term,
describing the action of the environment during the finite time that
the system Hamiltonian is switched on, has been given.

For further evaluation of ${\cal V}$ we adopt a matrix representation,
defined by
\begin{equation}
{\cal V}_{ab|cd}=(e_{ab},{\cal V} e_{cd})\equiv Tr_s e_{ab}^\dagger {\cal V}
e_{cd},
\end{equation}
where $\{e_{ab}| a,b=1,..,4\}$ is an orthonormal basis, i.e.
$(e_{ab},e_{cd})= \delta_{ac}\delta_{bd}$. In this notation we then
have
\begin{equation}
\rho(t)_{ab}=\sum_{c,d} {\cal V}_{ab|cd}(\rho_0)_{cd}
\end{equation}
with ${\cal V}$ being
a $16 \times 16$ matrix.

Note that ${\cal K}_{2,3}$ and ${\cal U}_s$ are not simultaneously
diagonal.  However, since ${\cal K}_3 (1,{\vec S}_i)=2\Gamma (0,{\vec
S}_i)$ we see that $\exp\{-(t-\tau_s){\cal K}_3\}$ is diagonal in the
``polarization basis" $\{e^{p}_{ab}=e_a^1 e_b^2 ;e_{1,..,4}^i=
(1/\sqrt{2},\sqrt{2}S^x_i,\sqrt{2}S^y_i,\sqrt{2}S^z_i), i=1,2\}$,
while ${\cal L}_s$ and thus ${\cal U}_s$ are diagonal in the
``multiplet basis" $\{e^m_{\alpha \beta}=|\alpha\rangle\langle
\beta|\, ,\alpha,\beta=1,...,4 \, ;\,
|1\rangle=(|01\rangle-|10\rangle)/\sqrt{2},
|2\rangle=(|01\rangle+|10\rangle)/\sqrt{2}, |3\rangle=|00\rangle,
|4\rangle=|11\rangle\}$, with
\begin{equation}
{\cal U}_s(t)_{\alpha\beta|\alpha'\beta'}=\delta_{\alpha\alpha'}
\delta_{\beta\beta'} e^{-it(E^m_\alpha-E^{m}_{\beta})},
\end{equation}
where $E^m_1=-3J_0/4$, $E^m_{2,3,4}=J_0/4$ are the singlet and triplet
eigenvalues.  Lastly, ${\cal K}_2$ is most easily evaluated also in
the multiplet basis; after some calculation we find that ${\cal
K}_2={\cal K}_2^d-{\cal K}_2^{nd}$, with
\begin{eqnarray}
({\cal
K}_2^d)_{\alpha\beta|\gamma\delta}&=&\sum_{i,\alpha'}[\delta_{\alpha\gamma}
\langle \delta|\vec S_i|\alpha'\rangle\cdot \langle\alpha'|\vec
S_i|\beta\rangle
k_{\alpha'\alpha'|\delta\beta}^*+\delta_{\beta\delta}
\langle \alpha|\vec S_i|\alpha'\rangle\cdot \langle\alpha'|\vec
S_i|\gamma\rangle
k_{\alpha'\alpha'|\gamma\alpha}],\\
({\cal K}_2^{nd})_{\alpha\beta|\gamma\delta}&=&\sum_i\langle\alpha|\vec
S_i|\gamma
\rangle\cdot\langle\delta|\vec S_i|\beta\rangle
\left(k_{\alpha\beta|\gamma\delta}+
(k_{\beta\alpha|\delta\gamma})^*\right).
\end{eqnarray}
Here
\begin{eqnarray}
\begin{array}{rll}
k_{\alpha\beta|\gamma\delta}&=&(\Gamma+i\Delta)e^{i(E^m_\delta-E^m_\beta)\tau_s}
\int_0^{\tau_s}d\tau e^{i(E^m_\alpha-E^m_\gamma)\tau}\\
\ &=&\frac{1}{2\omega_{\alpha\gamma}}[\Gamma c_{\delta\beta}-\Delta
s_{\delta\beta}
+i(\Gamma s_{\delta\beta}+\Delta c_{\delta\beta})][s_{\alpha\gamma}+
i(1-c_{\alpha\gamma})],\\
c_{ij}&=&\cos(\tau_s\omega_{ij}),\ \ s_{ij}=\sin(\tau_s\omega_{ij}),\ \
\omega_{ij}=
E^m_i-E^m_j.\end{array}\nonumber
\end{eqnarray}
Using the above matrix notation, we can write explicitly
\begin{equation}
{\cal V}_{\alpha\beta|\gamma\delta}=\sum_{a,b,\alpha',\beta'}
(C_{ab|\alpha\beta})^*
(e^{-(t-\tau_s){\cal K}_3})_{ab|ab}
C_{ab|\alpha'\beta'}
e^{-i\tau_s(E^m_{\alpha'}-E^m_{\beta'})}
(1-{\cal K}_2)_{\alpha'\beta'|\gamma\delta}\, ,
\end{equation}
where $C_{ab|\alpha\beta}=(e^p_{ab},e^m_{\alpha\beta})$ is the unitary
basis change between the polarization and the multiplet basis.

\subsection{One-bit gates}
\label{1b}

We now repeat the preceding analysis for single-qubit rotations such
as $e^{i {\pi\over 2}S_i^z }$ as required in Eq.~(\ref{XOR}).  Such
rotations can be achieved if a magnetic field ${\vec H}_i$ could be
pulsed exclusively onto spin $i$, perhaps by a scanning-probe tip.  An
alternative way, which would become attractive if further advances are
made in the synthesis of nanostructures in magnetic
semiconductors\cite{Awsch}, is to use, as indicated in Fig.~1(a), an
auxiliary dot (FM) made of an insulating, ferromagnetically-ordered
material that can be connected to dot 1 (or dot 2) by the same kind of
electrical gating as discussed above\cite{Harvard}.  If the the
barrier between dot 1 and dot FM were lowered so that the electron's
wavefunction overlaps with the magnetized region for a fixed time
$\tau_s$, the Hamiltonian for the qubit on dot 1 will contain a Zeeman
term during that time.  For all earlier and later times the magnetic
field seen by the qubit should be zero; any stray magnetic field from
the dot FM at neighboring dots 1, 2, etc. could be made small by
making FM part of a closure domain or closed magnetic circuit.

In either case, the spin is rotated and the corresponding Hamiltonian
is given by
\begin{equation}
\int_0^{\tau_s}dt\, H_s^H=\sum_{i=1}^2\omega_i\tau_s S_i^z,\label{hh}
\end{equation}
with $\omega_i=g \mu_B H_i^z$, where we assume that the H-field acting
on spin $i$ is along the {\em z}-axis.  The calculation proceeds along
the same line as the one described above: Just as in Eq. (\ref{k2}),
the expression obtained for the superoperator is
\begin{equation}
{\cal V}_H(t)=e^{-(t-\tau_s){\cal K}_3}{\cal U}^H_s(\tau_s)\,
(1-{\cal K}^H_2)\,\, .
\label{k2h}
\end{equation}
${\cal K}_3$ is exactly the same as before, Eq. (\ref{kthree}).
${\cal U}^H_s(\tau_s)$ is again given by Eq. (\ref{uuu}) with the
modification that the Liouvillian (see Eq. (\ref{lio})) corresponding
to the magnetic field Hamiltonian of Eq. (\ref{hh}) is used rather
than that for the exchange Hamiltonian $H_s$ (Eqs. (\ref{ham}) and
(\ref{tham})).  The explicit matrix representation is
\begin{equation}
({\cal U}^H_s(\tau_s))_{rs|r's'}=\delta_{rr'}
\delta_{ss'}e^{-i\sum_{i=1}^2(E^i_r-E^i_s)\tau_s}.
\end{equation}
Here we are employing another basis, the $S_z$-basis for the two spins
$\{e_{rs}^z=|r\rangle\langle s|,\ r,s=1,2,3,4;
|s\rangle=|00\rangle,|01\rangle,|10\rangle,|11\rangle\}$.  The
energies are
\begin{eqnarray}
\{E_r^1\}=\{E_{1,2,3,4}^1\}={{\omega_1}\over 2}\{1,1,-1,-1\},\nonumber \\
\{E_r^2\}=\{E_{1,2,3,4}^2\}={{\omega_2}\over 2}\{1,-1,1,-1\}.\label{eee}
\end{eqnarray}
The ${\cal K}^H_2$ calculation also proceeds as before (see
Eq. (\ref{ktwo})) using the new Hamiltonian; the result is ${\cal
K}^H_2={\cal K}_2^{H,d}-{\cal K}_2^{H,nd}$, with
\begin{eqnarray}
({\cal K}_2^{H,d})_{rs|tu}&=&\sum_{i,r'}\left(\delta_{rt}
\langle u|\vec S_i|r'\rangle\cdot \langle r'|\vec S_i|s\rangle
(k^i_{r'r'|us})^*+\delta_{su}
\langle r|\vec S_i|r'\rangle\cdot \langle r'|\vec S_i|t\rangle
k^i_{r'r'|tr}\right),\\
({\cal K}_2^{H,nd})_{rs|tu}&=&\sum_i\langle r|\vec S_i|t
\rangle\cdot\langle u|\vec S_i|s\rangle \left(k^i_{rs|tu}+
(k^i_{sr|ut})^*\right).
\end{eqnarray}
Here
\begin{eqnarray}
\begin{array}{rll}
k^i_{rs|tu}&=&(\Gamma+i\Delta)e^{i(E^i_u-E^i_s)\tau_s}
\int_0^{\tau_s}d\tau e^{i(E^i_r-E^i_t)\tau}\\
\ &=&\frac{1}{2\omega^i_{rt}}[\Gamma c^i_{us}-\Delta s^i_{us}
+i(\Gamma s^i_{us}+\Delta c^i_{us})][s^i_{rt}+
i(1-c^i_{rt})],\\
c^k_{ij}&=&\cos(\tau_s\omega^k_{ij}),\ \ s^k_{ij}=\sin(\tau_s\omega^k_{ij}),
\ \ \omega^k_{ij}=
E^k_i-E^k_j.\end{array}
\end{eqnarray}
The $E^k$'s are from Eq. (\ref{eee}).  Finally, the explicit matrix
form for ${\cal V}^H$ may be written
\begin{equation}
{\cal V}^H_{ab|a'b'}=\sum_{r,s,r',s'}(e^{-(t-\tau_s){\cal K}_3})_{ab|ab}
(D_{rs|ab})^*e^{-i\sum_{i=1}^2\tau_s(E^i_r-E^i_s)}
(1-{\cal K}^H_2)_{rs|r's'}D_{r's'|a'b'}\, ,
\end{equation}
where $D_{rs|ab}=(e^z_{rs},e^p_{ab})$ is now the unitary basis
change between the $S_z$ basis and the polarization basis.

\subsection{Numerical study for swap gate and XOR gate}

Having diagonalized the problem, we can now calculate any system
observable; the required matrix calculations are involved and complete
evaluation is done with Mathematica.  We will consider three
parameters (s, F, and P in Fig.~2) relevant for characterizing the
gate operation.  We first perform this analysis for the swap operation
introduced above.

The swap operation would provide a useful experimental test for the
gate functionality: Let us assume that at $t=0$ spin 2 is (nearly)
polarized, say, along the {\em z}-axis, while spin 1 is (nearly)
unpolarized, i.e. $\rho_0=(1+2S_2^z)/4$. This can be achieved, e.g.,
by selective optical excitation, or by an applied magnetic field with
a strong spatial gradient.  Next we apply a swap operation by pulsing
the exchange coupling such that $J_0\tau_s=\pi$, and observe the
resulting polarization of spin 1 described by
\begin{equation}
\langle S_1^z(t)\rangle={1\over 2}{\cal V}(t)_{41|14},
\label{spinswap}
\end{equation}
where ${\cal V}$ is evaluated in the polarization basis.  After time
$\tau_s$ spin 1 is almost fully polarized (whereas spin 2 is now
unpolarized) and, due to the environment, decays exponentially with
rate of order $\Gamma$.  To make the signal Eq. (\ref{spinswap})
easily measurable by conventional magnetometry, we can envisage a
set-up consisting of a large array of identical, non-interacting pairs
of dots as indicated in Fig.~1(b).

To further characterize the gate performance we follow
Ref. \cite{Poyatos} and calculate the gate fidelity $F= {\overline
{\langle \psi_0|{\cal U}^\dagger(\tau_s)\rho(t)|\psi_0\rangle}}$, and
the gate purity $P={\overline {Tr_s[\rho(t)]^2}}$, where the overbar
means average over all initial system states $|\psi_0\rangle$.
Expressing ${\cal V}$ in the multiplet basis and using trace and
Hermiticity preservation we find
\begin{eqnarray}
F(t)&=&{1\over 6} +{1\over 24} {\rm Re}[\sum_\alpha {\cal
V}_{\alpha\alpha|\alpha\alpha}
+\sum_{\alpha,\beta} {\cal V}_{\alpha\beta|\alpha\beta}
e^{i\tau_s(E^m_\alpha-E^m_\beta)}],
\\ P(t)&=&{1\over 24}\sum_{i,k,k'}[|{\cal V}_{kk'|ii}|^2+\sum_{j}(
{\cal V}_{kk'|ii}{\cal V}^*_{kk'|jj}+|{\cal V}_{kk'|ij}|^2)]
\end{eqnarray}
(in fact, the expression for $P(t)$ holds in any basis).  Evaluations
of these functions for specific parameter values are shown in Fig.\ 2.
For the parameters shown, the effect of the environment during the
switching, i.e. ${\cal K}_2$ in Eq.~(\ref{k2}), is on the order of a
few percent.

The dimensionless parameters used here would, for example, correspond
to the following actual physical parameters: if an exchange constant
$J_0= 80\mu {\rm eV}\approx 1K$ were achievable, then pulse durations
of $\tau_s\approx 25{\rm ps}$ and decoherence times of
$\Gamma^{-1}\approx 1.4{\rm ns}$ would be needed; such parameters, and
perhaps much better, are apparently achievable in solid-state spin
systems\cite{Awsch}.

As a final application, we calculate the full XOR by applying the
corresponding superoperators in the sequence associated with the one
on the r.h.s. of Eq. (\ref{XOR}).  We use the same dimensionless
parameters as above, and as before we then calculate the gate fidelity
and the gate purity.  Some representative results of this calculation
are plotted in the inset of Fig.~2(b).  To attain the $\pi/2$
single-bit rotations of Eq. (\ref{XOR}) in a $\tau_s$ of $25{\rm ps}$
would require a magnetic field $H\approx 0.6T$, which would be readily
available in the solid state.

\section{Discussion}

As a final remark about the decoherence problem, we would note that
the parameters which we have chosen in the presentation of our
numerical work, which we consider to be realistic for known nanoscale
semiconductor materials, of course fall far, far lower than the
0.99999 levels which are presently considered desirable by
quantum-computation theorists\cite{BD}; still, the achievement of even
much lesser quality quantum gate operation would be a tremendous
advance in the controlled, non-equilibrium time-evolution of
solid-state spin systems, and could point the way to the devices which
could ultimately be used in a quantum computer.

Considering the situation more broadly, we are quite aware that our
proposal for quantum dot quantum computation relies on simultaneous
further advances in the experimental techniques of semiconductor
nano-fabrication, magnetic semiconductor synthesis, single
electronics, and perhaps in scanning-probe techniques.  Still, we also
strongly feel that such proposals should be developed seriously, and
taken seriously, in the present, since we believe that many aspects of
the present proposal are testable in the not-too-distant future. This
is particularly so for the demonstration of the swap action on an
array of dot pairs.  Such a demonstration would be of clear interest
not only for quantum computation, but would also represent a new
technique for exploring the non-equilibrium dynamics of spins in
quantum dots.

To make the quantum-dot idea a complete proposal for quantum
computation, we need to touch on several other important features of
quantum-computer operation.  As our guideline we follow the five
requirements laid out by one of us\cite{DDV}: 1) identification of
well-defined qubits; 2) reliable state preparation; 3) low
decoherence; 4) accurate quantum gate operations; 5) strong quantum
measurements.  Items 1, 3, and 4 have been very thoroughly considered
above.  We would now like to propose several possible means by which
requirements 2 and 5, for state preparation (read in) and quantum
measurement (read out), may be satisfied.

One scheme for qubit measurement which we suggest involves a
switchable tunneling (T in Fig.~1(a)) into a supercooled paramagnetic
dot (PM).  When the measurement is to be performed, the electron
tunnels (this will be real tunneling, not the virtual tunneling used
for the swap gate above) into PM, nucleating from the metastable phase
a ferromagnetic domain whose magnetization direction could be measured
by conventional means.  The orientation $(\theta,\phi)$ of this
magnetization vector is a ``pointer'' which measures the spin
direction; it is a generalized measurement in which the measurement
outcomes form a continuous set rather than having two discrete values.
Such a case is covered by the general formalism of
positive-operator-valued (POV) measurements\cite{Peres}.  If there is
no magnetic anisotropy in dot PM, then symmetry dictates that the
positive measurement operators would be projectors into the
over-complete set of spin-1/2 coherent states
\begin{equation}
|\theta,\phi\rangle=\cos{\theta\over 2}|0\rangle+e^{i\phi}\sin{\theta\over
2}|1\rangle.
\end{equation}
A 75\%-reliable measurement of spin-up and spin-down is obtained if
the magnetization direction $(\theta,\phi)$ in the upper hemisphere is
interpreted as up, and in the lower hemisphere as down; this is so
simply because
\begin{equation}
{1\over{2\pi}}\int_Ud\Omega|\langle 0|\theta,\phi\rangle|^2={3\over 4}.
\end{equation}
Here ``$U$" denotes integration over the upper hemisphere and $2\pi$
is the normalization constant for the coherent states.

Another approach which would potentially give a 100\% reliable
measurement requires a spin-dependent, switchable ``spin valve''
tunnel barrier (SV) of the type mentioned e.g. in Ref.  \cite{Prinz}.
When the measurement is to be performed, SV is switched so that only
an up-spin electron passes into semiconductor dot 3.  Then the
presence of an electron on 3, measured by electrometer ${\cal E}$,
would provide a measurement that the spin had been ``up.''  It is well
known now how to create nanoscale single-electron electrometers with
exquisite sensitivity (down to $10^{-8}$ of one
electron)\cite{Devoret}.

We need only discuss the state-preparation problem briefly.  For many
applications in quantum computing, only a simple initial state, such
as all spins up, needs to be created.  Obviously, such a state is
achieved if the system is cooled sufficiently in a uniform applied
magnetic field; acceptable spin polarizations of electron spins are
readily achievable at cryogenic temperatures.  If a specific
arrangement of up and down spins were needed as the starting state,
these could be created by a suitable application of the reverse of the
spin-valve measurement apparatus.

\acknowledgments We are grateful to D. D. Awschalom, H.-B. Braun,
T. Brun and G. Burkard for useful discussions. This research was
supported in part by the National Science Foundation under Grant
No. PHY94-07194.

\appendix
\section{Complete positivity of time-evolution superoperator $\cal V$}

Here we sketch the proof that the superoperator $\cal V$ in
Eq. (\ref{k2}) is completely positive.  We analyze the ${\cal K}_3$
term first.  We write
\begin{equation}
e^{-\tau{\cal K}_3}=\lim_{N\rightarrow\infty}\left(1-{\tau\over N}{\cal
K}_3\right)^N.
\end{equation}
It is sufficient to prove that the infinitesimal operator is
completely positive.  It is straightforward to show, using
Eq. (\ref{kthree}), that
\begin{equation}
\left(1-{\tau\over N}{\cal K}_3\right)\rho=Z_3^\dagger\cdot\rho Z_3+O\left(
(\tau/N)^2\right).
\label{Kraus}
\end{equation}
Here $Z_3$ is the seven-component vector operator
\begin{equation}
Z_3=\left (1-{\tau\over{2N}}\sum_{k=1}^6B_k^\dagger B_k\ ,\sqrt{\tau\over
N}{\bf B}\right),
\end{equation}
where
\begin{equation}
{\bf B}=(B_1,...,B_6)=\sqrt{2\Gamma}\,\,({\vec S}_1,{\vec S}_2).
\end{equation}
Note that for this case $B^\dagger_k=B_k$ and $\sum_{k=1}^6B_k^\dagger
B_k=3 \Gamma$.

We recall that it is easy to prove that any superoperator $\cal S$ of
the form
\begin{equation}
{\cal S}\rho=Z^\dagger\rho Z\label{genKraus}
\end{equation}
as in the first term of Eq. (\ref{Kraus}) is completely positive.
Indeed, considering its action on any state vector of the system plus
environment $\phi$, and taking a positive $\rho$ we get
\begin{equation}
(\phi,{\cal S}\rho\phi)=(\phi,Z^\dagger\rho Z\phi)=(Z\phi,\rho Z\phi)\ge 0
\ \ \forall\phi.
\end{equation}

Next we consider the $1-{\cal K}_2$ term of Eq. (\ref{k2}).  Starting
from Eq. (\ref{ktwo}), we put this term in a form corresponding to the
completely-positive form Eq. (\ref{genKraus}).  We find
\begin{equation}
(1-{\cal K}_2)\rho=Z_2^\dagger\cdot\rho Z_2+O\left(\lambda^4,\tau_s^2,
(\lambda^2\tau_s)^2\right),
\label{k2almost}
\end{equation}
with $Z_2$ being the vector operator
\begin{equation}
Z_2=(1+Y^\dagger\cdot X^\dagger,X-Y^\dagger)
\end{equation}
with
\begin{eqnarray}
X&=&-(\Gamma+i\Delta)\left(\vec S_1(\tau_s),\vec S_2(\tau_s)\right),\\
Y&=&\int_0^{\tau_s}d\tau\left(\vec S_1(\tau),\vec S_2(\tau)\right).
\end{eqnarray}
So, from the same arguments as above, Eq. (\ref{k2almost}) establishes
that $1-{\cal K}_2$ is completely positive up to the order of accuracy
discussed in the text.

Finally, we note that the other two general conditions for a physical
superoperator also follow immediately: Trace preservation of $\cal V$
follows from the fact that a Liouvillian $\cal L$ appears to the left
in the basic equations for ${\cal K}_2$, Eq. (\ref{k2lio}), and ${\cal
K}_3$, Eq. (\ref{k3lio}).  Trace preservation is also reflected in the
fact that $Z_2\cdot Z_2^\dagger=1$ and $Z_3\cdot Z_3^\dagger=1$ to
leading order.  The form Eq. (\ref{genKraus}) also obviously preserves
Hermiticity of the density operator; this is also clear from the forms
of Eqs. (\ref{ktwo}) and (\ref{kthree}).

\begin{figure}
\epsfxsize=15cm
\leavevmode
\epsfbox{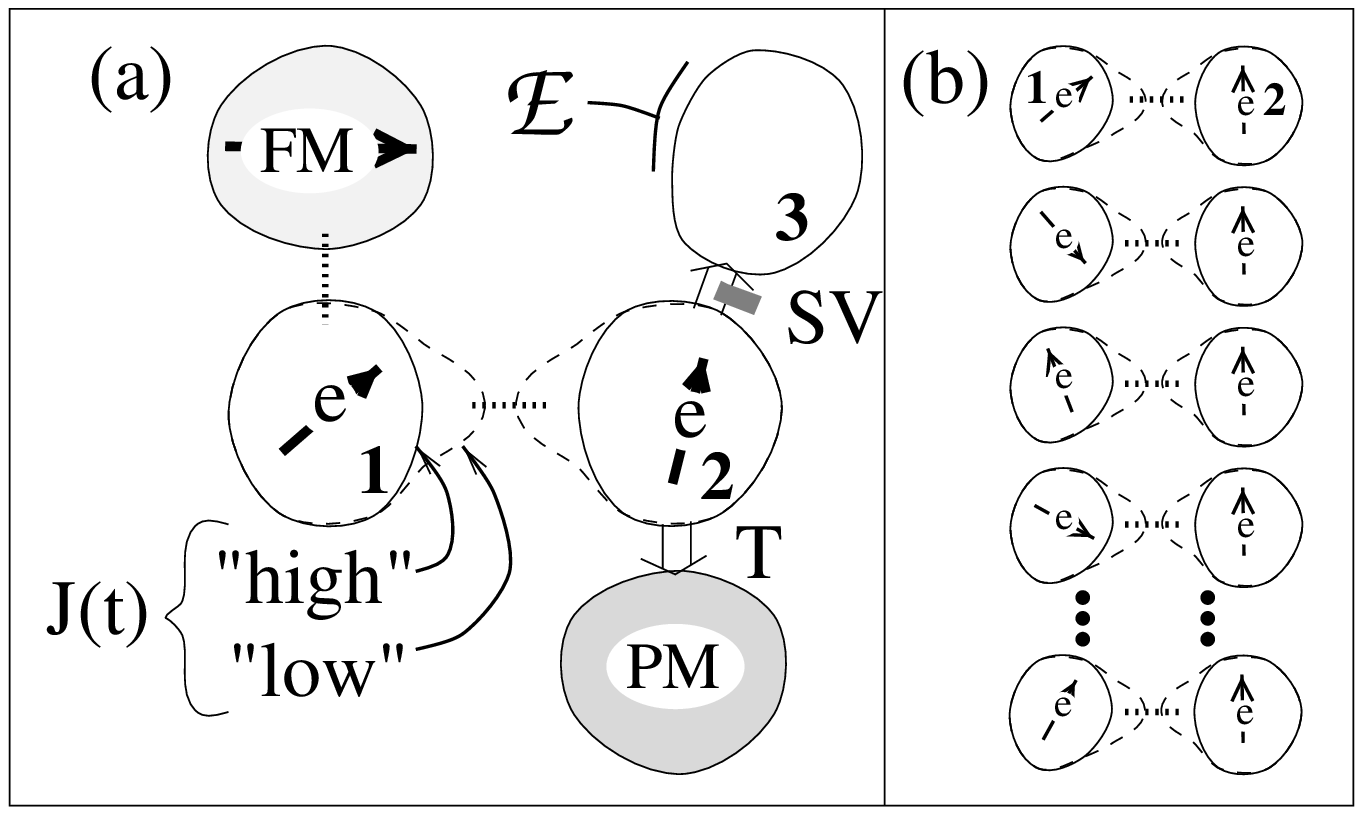}
\caption{ a) Schematic top view of two coupled quantum dots labeled 1
and 2, each containing one single excess electron (e) with spin
1/2. The tunnel barrier between the dots can be raised or lowered by
setting a gate voltage ``high'' (solid equipotential contour) or
``low'' (dashed equipotential contour).  In the low state virtual
tunneling (dotted line) produces a time-dependent Heisenberg exchange
$J(t)$.  Hopping to an auxiliary ferromagnetic dot (FM) provides one
method of performing single-qubit operations.  Tunneling (T) to the
paramagnetic dot (PM) can be used as a POV read out with 75\%
reliability; spin-dependent tunneling (through ``spin valve'' SV) into
dot 3 can lead to spin measurement via an electrometer ${\cal E}$.  b)
Proposed experimental setup for initial test of swap-gate operation in
an array of many non-interacting quantum-dot pairs.  Left column of
dots is initially unpolarized while right one is polarized; this state
can be reversed by a swap operation (see Eq.\
(\protect{\ref{spinswap}})).}
\label{model}
\end{figure}

\begin{figure}
\epsfxsize=15cm
\leavevmode
\epsfbox{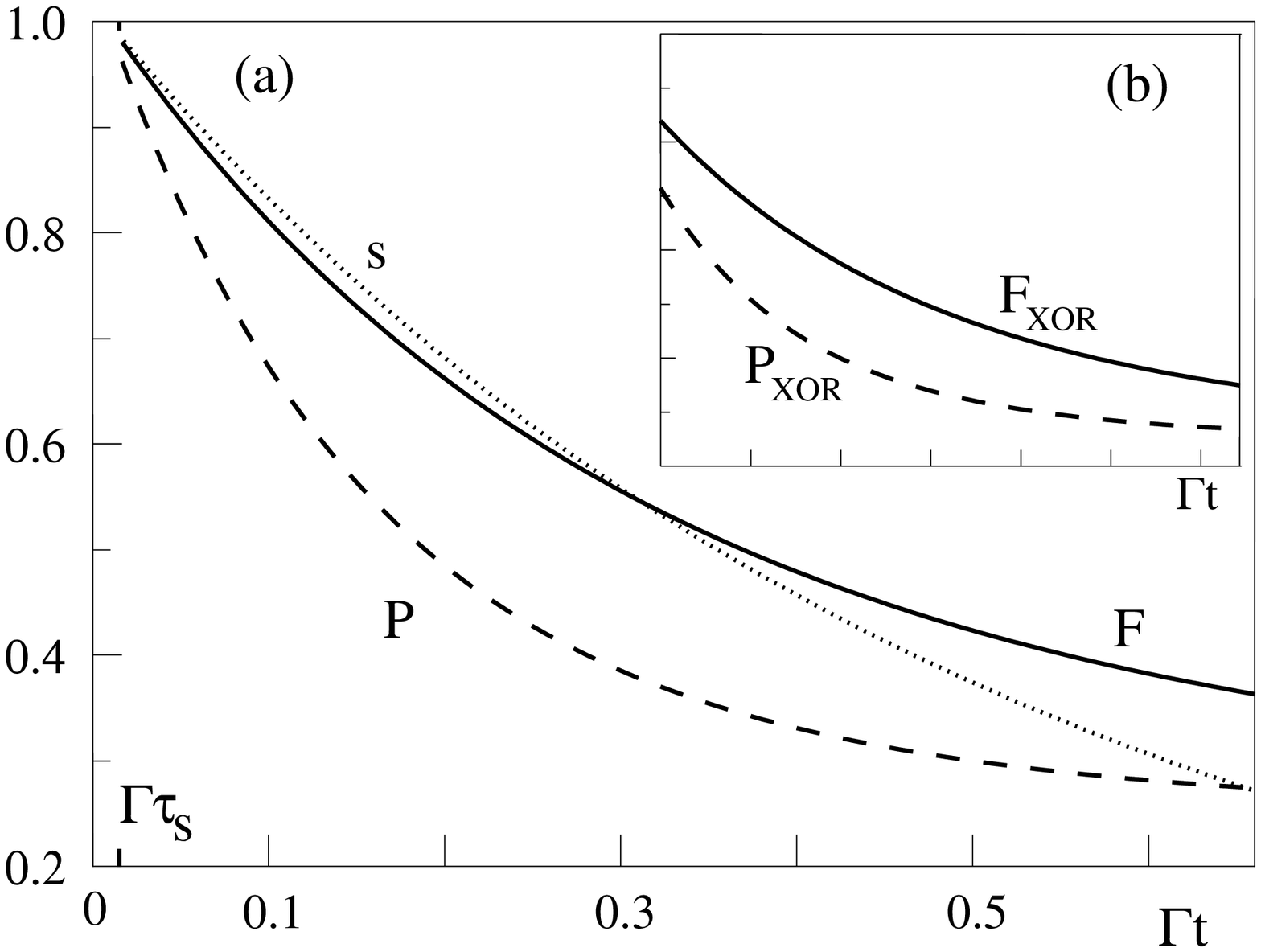}
\caption
{ a) Swap polarization ${\rm s}\equiv2\langle S_1^z(t)\rangle$ (see
Eq.~(\protect\ref{spinswap})), gate fidelity F, and gate purity P {\it
vs.}\ $\Gamma t$ for ``swap'' using parameters $J_0\tau_s=\pi$,
$\Gamma\tau_s=0.017$, and $\Delta\tau_s=-0.0145$. b) The same for XOR
obtained using the four operations in Eq.\ (\protect{\ref{XOR}}) (the
final two single-spin operations done simultaneously).  The same
parameters and scales as in (a) are used; the pulse-to-pulse time is
taken to be $3\tau_s$.  $\Gamma t$ is measured from the end of the
fourth pulse.}
\label{plots}
\end{figure}


\begin{references}
\bibitem{superexchange}
We can also envisage a superexchange mechanism to obtain a Heisenberg
interaction by using three aligned quantum dots where the middle one
has a higher energy level (by the amount $\epsilon$) such that
the electron spins of the outer
two dots are also Heisenberg coupled, but now with the exchange
coupling being $J=4t_0^4(1/\epsilon^2 u + 1/2\epsilon^3)$.
\bibitem{pulseform} We assume for simplicity that the shape of
the applied pulse is roughly rectangular with $J_0 \tau_s$ constant.
\bibitem{footnoteHXOR} We note that explicitly
$U_{XOR}={1\over 2}+S_1^z +S_2^z- 2S_1^z S_2^z$,
with the corresponding XOR Hamiltonian,
$\int_0^t dt' H_{XOR}= \pi[1 -2S_1^z -2S_2^z + 4S_1^z S_2^z]/4$.
An alternative way to achieve the XOR operation is
given by $U_{XOR}=
e^{i {\pi}S_1^z }U_{sw}^{-\frac{1}{2}}
e^{-i {\pi\over 2}S_1^z}
U_{sw}e^{i{\pi\over 2} S_1^z}U_{sw}^\frac{1}{2}$.
This form has the potential advantage that the single qubit operations
involve only spin 1.
\end{references}
\end{document}